\documentclass[prl,twocolumn,showpacs,amsmath,amssymb]{revtex4}

\usepackage{color,epsfig,rotating,graphicx,subfigure,ulem}
\usepackage{amsmath,amssymb,amscd,comment}
\usepackage{mathrsfs}
\usepackage[latin1]{inputenc}
\usepackage[english]{babel}

\usepackage{dsfont}
\usepackage{textcomp}

\newcommand{\rd}{{\rm d}}

\newcommand{\kb}{k_{\rm B}}

\newcommand{\kB}{k_{\rm B}}

\begin{document}

\title{Non-trivial topological structure of heat and momentum flux radiated by magneto-optical nanoparticles.}
\author{A. Ott$^{1}$}
\author{P. Ben-Abdallah$^{2}$}
\email{pba@institutoptique.fr}
\author{S.-A. Biehs$^{1}$}
\email{s.age.biehs@uni-oldenburg.de}
\affiliation{$^1$ Institut f\"{u}r Physik, Carl von Ossietzky Universit\"{a}t, D-26111 Oldenburg, Germany}
\affiliation{$^2$ Laboratoire Charles Fabry,UMR 8501, Institut d'Optique, CNRS, Universit\'{e} Paris-Sud 11,
2, Avenue Augustin Fresnel, 91127 Palaiseau Cedex, France}

%
\begin{abstract}
In the present Letter we investigate the heat and momentum fluxes radiated by a hot magneto-optical nanoparticle in its surrounding under the action of an external magnetic field.  We show that the flux lines circulate in a confined region at nanometric distance from the particle around the axis of magnetic field  in a vortex-like configuration. Moreover we prove that the spatial orientation of these vortices (clockwise or counterclockwise) is associated to the contribution of optical resonances with topological charges $m = +1$ or $m = -1$  to the thermal emission.  This work paves the way to a geometric description of  heat and momentum transport in lattices of magneto-optical particles. Moreover it could have important applications in the field of energy storage as well as in thermal management at nanoscale.  
\end{abstract}

\maketitle


Unconventional topological magnetic textures such as magnetic spin vortex (magnetic skyrmions~\cite{Muhlbaue,Neubauer,Pappas})  has given rise to radically new effects in solids state physics such as the spin Hall effect~\cite{Chen}). In particular, these topological defects have found  important applications in spintronics~\cite{Fert} by allowing for the  developpment of novel devices for information storage or logic treatment at nanoscale. 
In this Letter we investigate a thermal analog of these topological defects obtained by texturing the heat and momentum flux around hot bodies when they radiate in their surrounding environment. We show that magneto-optical (MO) nanoparticles in out of thermal equilibrium situation give rise, under the presence of an external magnetic field, to swirling structures of heat and momentum fluxes. Moreover we demonstrate that the topological number associated to these structures can be tuned either by changing the magnitude of magnetic field or by heating/cooling down the particle. Finally we infer that this geometric behavior could explain several thermo-magnetic effects such as the thermal radiative Hall effect, persistent currents, and giant magneto-resistance~\cite{Ben-Abdallah2016,Zhu2016,Latella2017,Cuevas}

\begin{figure}
	\centering
        \includegraphics[width=0.45\textwidth]{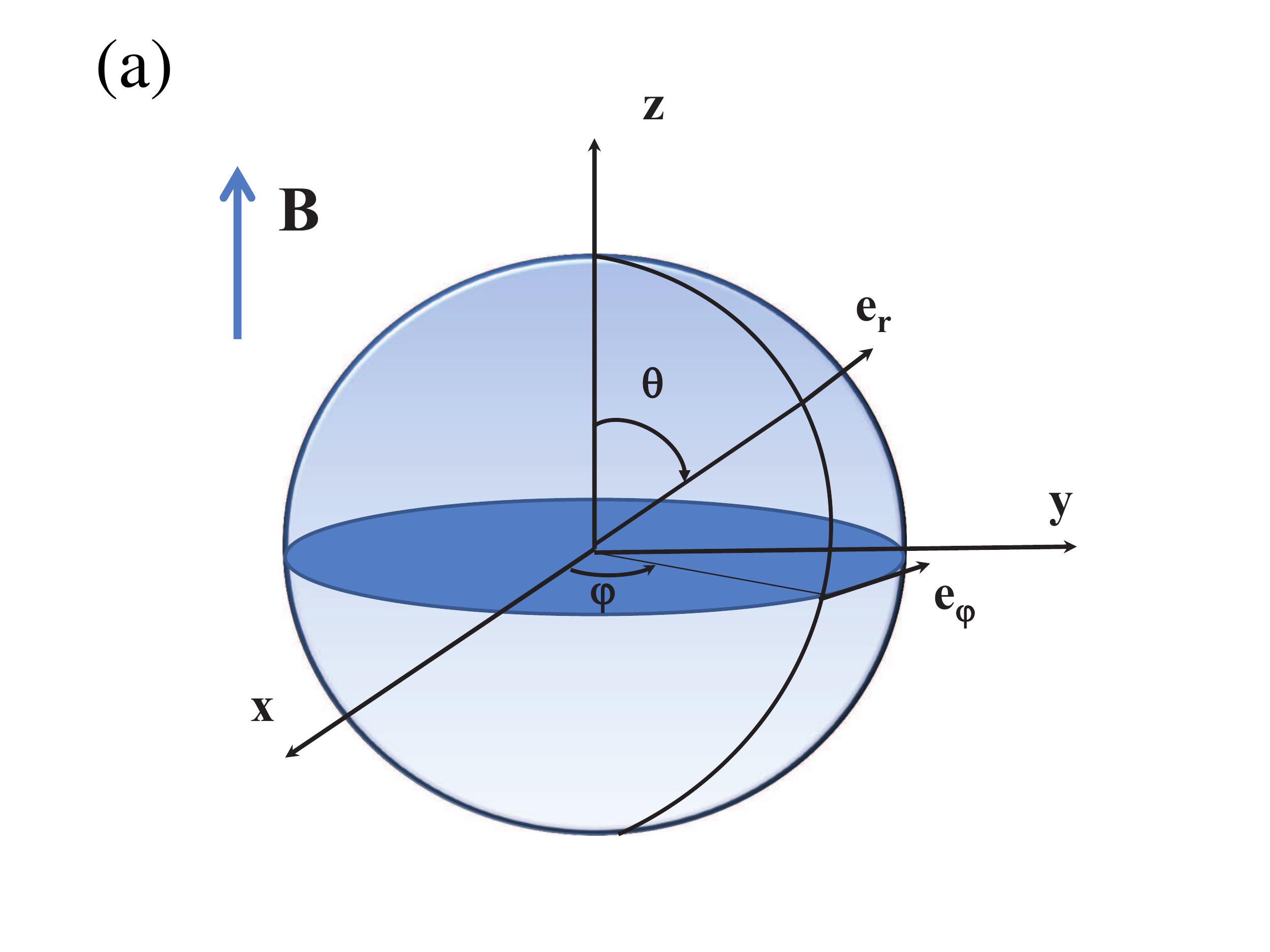}
	\includegraphics[width=0.45\textwidth]{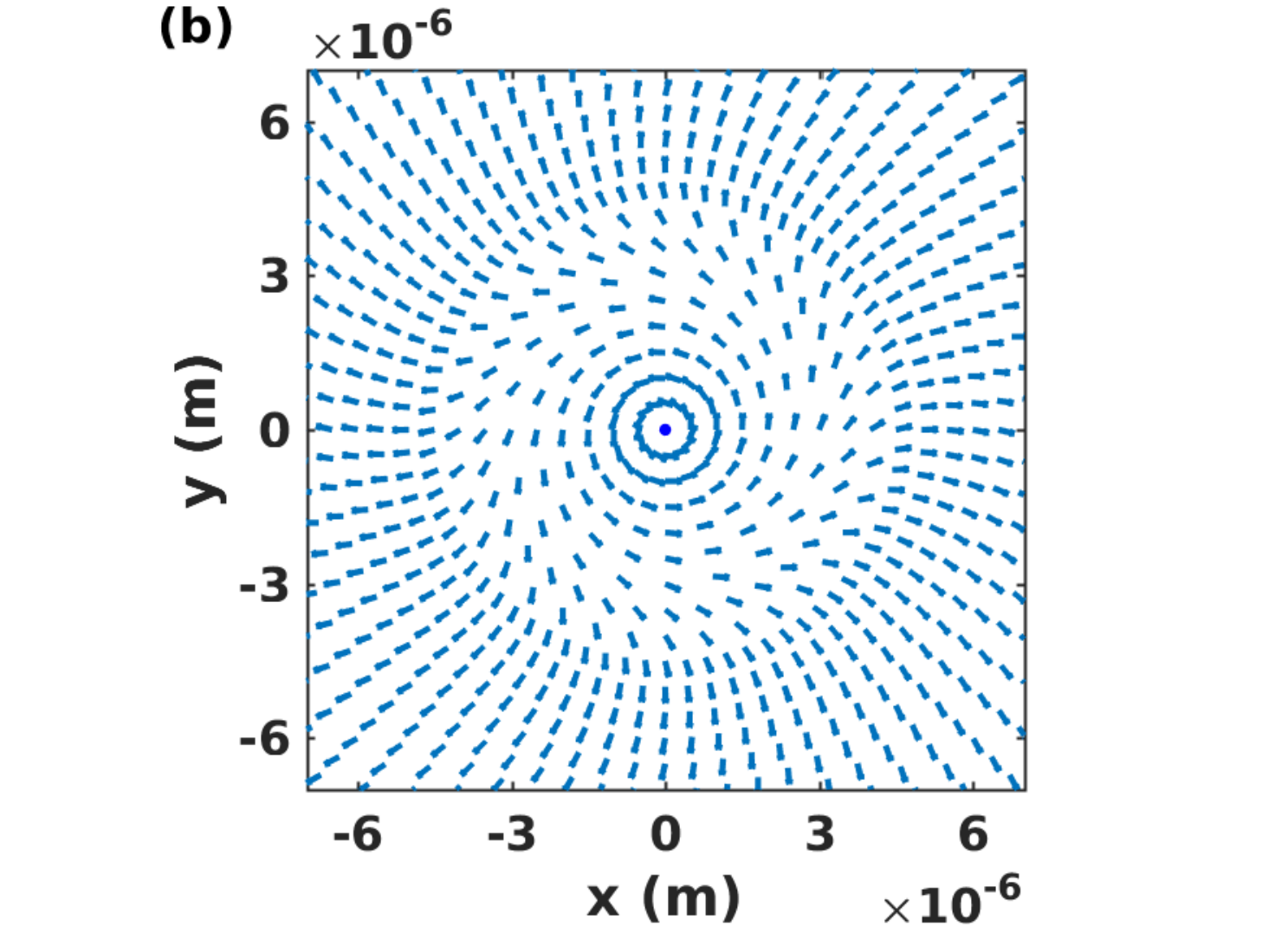}
	\caption{(a) Sketch of MO particle and coordinate system. (b) Mean PV around a InSb nanoparticle 100~nm radius at temperature T=300 K for a magnetic field of 5T  plotted as a vector field  in the x-y plane and
 showing the transition between the near-field and far-field regime around a distance of about 6 microns.}
	\label{ebenen}
\end{figure}

To start, we consider the system sketched in Fig.~1(a). It consists in a spherical nanoparticle of radius $R$ made of a MO material. The  particle is supposed to be held at temperature $T$, immersed in a transparent surrounding medium at temperature $T_a = 0\,{\rm K}$ and subjected to a constant magnetic field $\vec{B} = B \vec{e}_z$ pointing in z direction. The thermal field radiated by this particle can generally be determined in the framework of Rytov's fluctuational electrodynamics~\cite{Rytov1989} by modelling the nanoparticle as a point dipole with a thermally fluctuating dipole moment $\vec{p}$
which has a zero mean value, i.e.\ $\langle \vec{p} \rangle = \vec{0}$. The correlation function can be determined by
the fluctuation-dissipation theorem giving
\begin{equation}
  \langle \vec{p} \otimes \vec{p}^*\rangle = \frac{2\epsilon_0}{\omega} \Theta(\omega,T) \frac{1}{2i}(\underline{\underline{\rm \alpha}}-\underline{\underline{\rm \alpha}}^\dagger)
\label{dipolmoment}
\end{equation}
with the mean energy of a harmonic oscillator $ \Theta(\omega,T) = \hbar\omega/\bigl[\exp(\hbar\omega/\kb T)-1\bigr]$; $\kB$ and $\hbar$ stand for the Boltzmann and reduced Planck constant. Here the matrix $\uuline{\alpha}$ stands for the polarizability of the nanoparticle. For an anisotropic particle the quasi-static expression has the form~\cite{Albaladejo}
\begin{equation}
  \underline{\underline{\rm \alpha}} = 4\pi R^3(\underline{\underline{\rm \epsilon}}-\mathds{1})(\underline{\underline{\rm \epsilon}}+2\mathds{1})^{-1}.
\label{alpha}
\end{equation}
The permittivity tensor $\uuline{\epsilon}$ describes as usual the optical properties of the nanoparticle. For a MO nanoparticle with
an externally applied magnetic field in z-direction it has the form
\begin{eqnarray}
  \underline{\underline{\rm \epsilon}}=\begin{pmatrix} \epsilon_1 & -i\epsilon_2 & 0 \\ i\epsilon_2 & \epsilon_1 & 0 \\ 0 & 0 & \epsilon_3 \end{pmatrix}. 
\label{epsilon}
\end{eqnarray}
In particular, for InSb we can explicitely express the components of the permittivity tensor as~\cite{Moncada}
\begin{align}
  \epsilon_1 &= \epsilon_\infty\left(1+\frac{\omega_L^2-\omega_T^2}{\omega_T^2-\omega^2-i\Gamma\omega}+\frac{\omega_p^2(\omega+i\gamma)}{\omega[\omega_c^2-(\omega+i\gamma)^2]} \right), \\
  \epsilon_2 &= \frac{\epsilon_\infty\omega_p^2\omega_c}{\omega[(\omega+i\gamma)^2-\omega_c^2]}, \\
  \epsilon_3 &= \epsilon_\infty\left(1+\frac{\omega_L^2-\omega_T^2}{\omega_T^2-\omega^2-i\Gamma\omega}-\frac{\omega_p^2}{\omega(\omega+i\gamma)^2} \right),
\end{align}
where  the material parameters are $\epsilon_\infty = 15.7$,   $\omega_L = 3.62\cdot 10^{13}\,{\rm rad/s}$, $\omega_T = 3,39\cdot10^{13}\,{\rm rad/s}$, $n = 1,07\cdot10^{17}\,{\rm cm}^{-3}$, $m^* = 1,99\cdot10^{32}\,{\rm kg}$, $\omega_p = \sqrt{\frac{ne^2}{m^*\epsilon_0\epsilon_\infty}}$, $\Gamma = 5,65\cdot10^{11}\,{\rm rad/s}$, $\gamma = 3,39\cdot10^{12}\,{\rm rad/s}$, and $\omega_c = \frac{eB}{m^*}$.

The spectral components of the electric and magnetic field of a an oscillating dipole with dipole moment $\mathbf{p}$ 
can be written as~\cite{Novotny} $ \vec{E} = \omega^2\mu_0\mathds{ G}^E(\vec{r},\vec{r'},\omega)\cdot\vec{p}(\omega,t)$ and $\vec{H} = \omega^2 \mu_0 \mathds{ G}^H(\vec{r},\vec{r'},\omega)\cdot\vec{p}(\omega,t)$,
where the electric and the magnetic Green functions are defined as~\cite{Novotny}
\begin{align}
  \mathds{G}^{E}(\vec{r},\vec{r'}) &= \frac{e^{ik_0r}}{4\pi r}\left[a\mathds{1}+b\vec{e_r}\otimes\vec{e_r}\right] , \\
  \mathds{G}^H(\vec{r},\vec{r'})   &= \sqrt{\frac{\epsilon_0}{\mu_0}}\frac{e^{ik_0r}}{4\pi r}l (\vec{e}_{\phi}\otimes\vec{e}_{\theta}-\vec{e}_{\theta}\otimes\vec{e}_{\phi}).
\end{align}
Here we have introduced the quantities $k_0 = \omega/c$ and
\begin{equation}
  a = 1+\frac{ik_0r-1}{k_0^2r^2}, b =\frac{3-3ik_0r-k_0^2r^2}{k_0^2r^2}, l =1+\frac{i}{k_0r}.
\end{equation}
With these expressions we can evaluate the mean Poynting vector (PV)
\begin{equation}
  \langle \vec{S} \rangle = \int_0^\infty \!\!\! \frac{\rd \omega}{2 \pi} \vec{S}_{\omega} = \int_0^\infty \!\!\! \frac{\rd \omega}{2 \pi}  2{\rm Re}\langle(\vec{E}_{\omega}\times\vec{H}^*_{\omega})\rangle.
\end{equation}
After a straighforward calculation the spectral PV in spherical coordinates reads
\begin{equation}
\begin{split}
  \vec{S}_\omega &=   \frac{\Theta(\omega,T) k_0^3}{4\pi^2r^2} \bigl[ \bigl(2 \alpha_{11}''+ (\alpha_{33}'' - \alpha_{11}'')\sin^2(\theta)\bigr) \vec{e}_r \\
                 &\quad + \alpha_{12}'\biggl(\frac{1}{k_0r}+\frac{1}{k_0^3r^3}\biggr)\sin(\theta) \vec{e}_\varphi\bigr].
\end{split}
\end{equation}
Due to the rotational symmetry around the magnetic field axis the PV does not depend on the azimutal angle $\varphi$. But interestingly the anisotropy of the nanoparticles generates a component of PV in the $\varphi$-direction introducing so a twist. For an isotropic nanoparticle $ S_{\phi,\omega} $ would be zero. From the above expression it is also clear that the heat flux orientation in $\pm \varphi$ direction depends on the sign of $ \alpha_{12}'$. This circular heat flux is presented in Fig.~\ref{ebenen}(b) where we have plotted the full mean PV in the x-y plane. It can be seen that the  heat flux is circular around the nanoparticle. Interestingly, we see that in the near-field regime the PV is counterclockwise whereas in the far-field regime it is clockwise oriented. This topological structure or local winding of the heat flux is at the origin of the observed symmetry breaking in the thermal Hall effect~\cite{Ben-Abdallah2016} and the persistent current~\cite{Zhu2016}.

\begin{figure}
	\centering
	\includegraphics[width=0.45\textwidth]{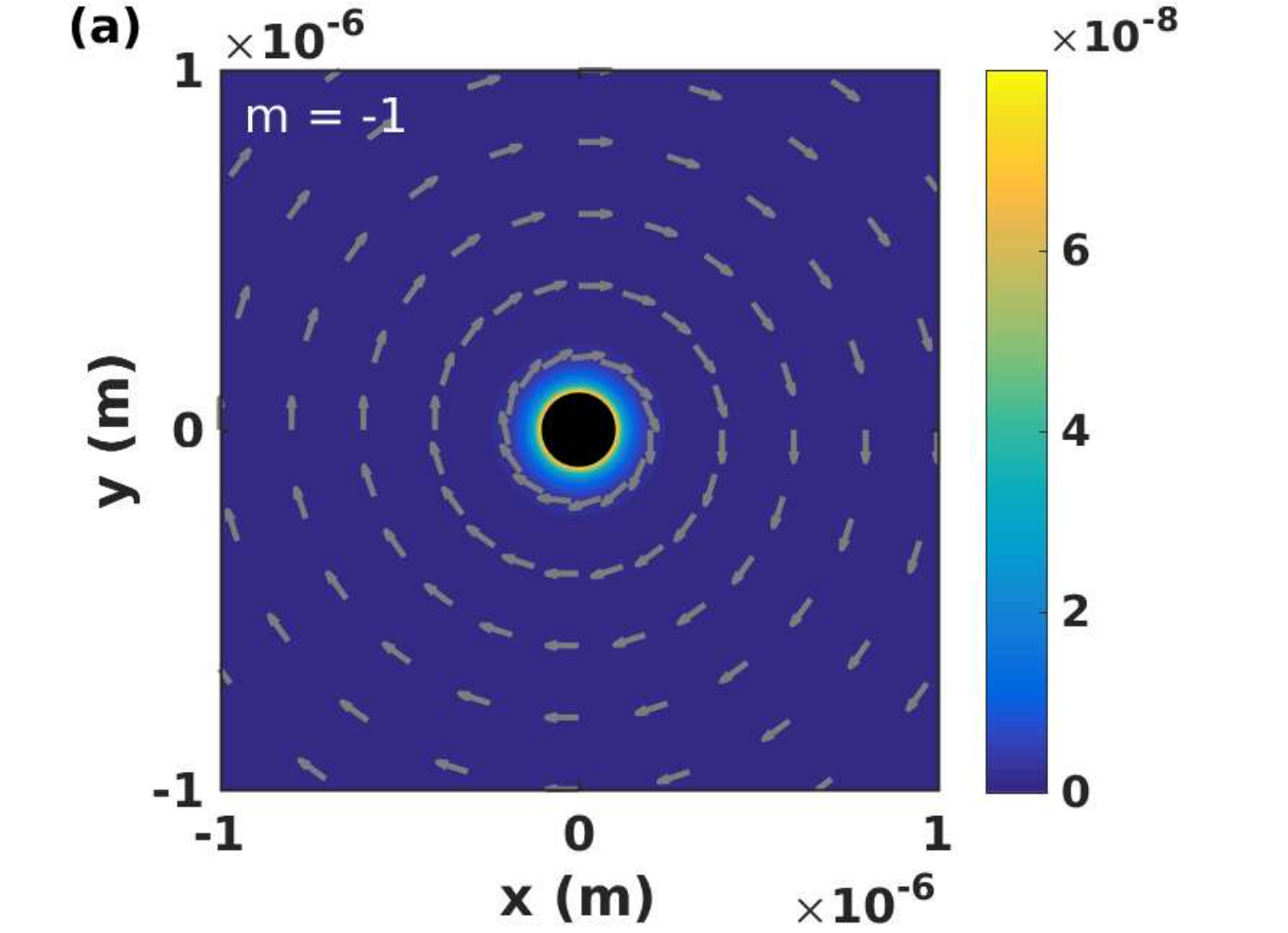}
	\includegraphics[width=0.45\textwidth]{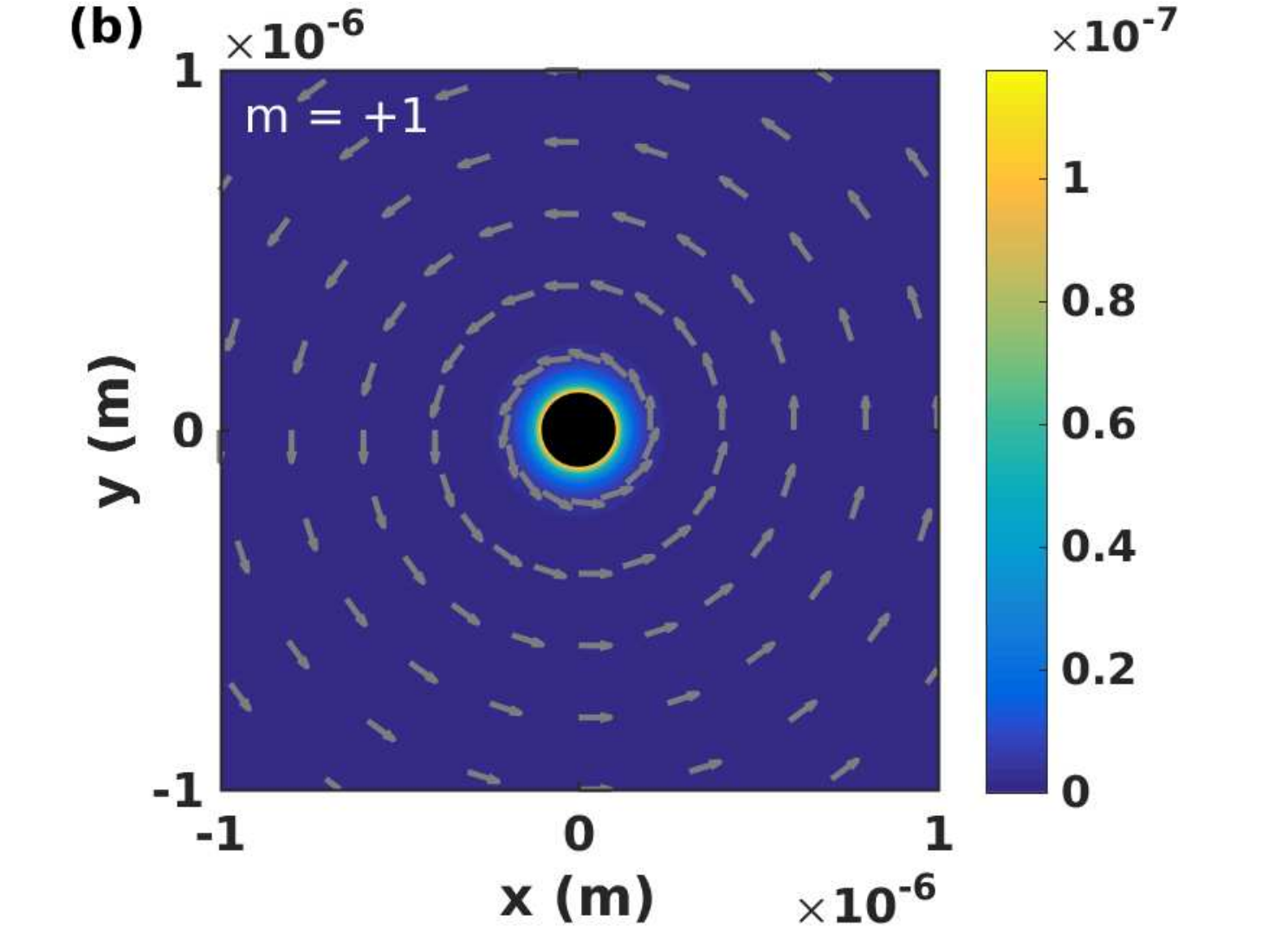}
	\caption{Spectral PV $\vec{S}_{\omega}$ as a normalised vector field in x-y plane together with the modulus of the spectral PV in Ws/m$^2$ for a magnetic field of 1T; $R$ = 100nm. In (a) we chose the resonance frequency $\omega_{m = -1}$ of the mode with $m = -1$ and in (b) we chose the resonance frequency $\omega_{m = +1}$ of the mode with $m = +1$.}
	\label{poyntingwres}
\end{figure}

This effect is connected to the dipolar eigenmodes of the nanoparticle which are determined of the poles of the polarizability that is by the roots of the following equation
\begin{equation}
\det \left( \underline{\underline{\rm \epsilon}}+2\mathds{1} \right) = (\epsilon_3 + 2)
\left[ (\epsilon_1 + 2)^2 - \epsilon^2_2 \right]=0.
\end{equation}
When no external magnetic field is applied $\epsilon_1 = \epsilon_3$ and $\epsilon_2 = 0$ so that the dipolar resonant modes with topological charges m = -1,0,1 are degenerate and their resonance frequency is solution of the usual equation
\begin{equation}
  \epsilon_3(\omega_{m = -1,0,1}) = -2.
\end{equation}
On the other hand, when the field is turned on the degeneracy between the two modes with topological 
charges $m = -1$ and $m = +1$ is lifted and the resonance frequency associated to these modes are obtained by solving the following equation
\begin{equation}
  \epsilon_1(\omega_{m = \pm 1}) = - 2 \mp \epsilon_2. 
\end{equation}
These two helical modes are connected with a circular motion of the electrons in the nanoparticle in $\pm \varphi$ direction~\cite{PineiderEtAl2013} which results in a circular heat flux similar to the optical vortices in paraxial beams~\cite{SoskinEtAl} with angular momentum $l =1$ and
$m = \pm 1$. This fact can be seen nicely in Fig.~\ref{poyntingwres} where we have plotted the spectral heat flux in the x-y plane
for the two frequencies $\omega_{m = \pm 1}$. 

\begin{figure}
	\centering
        \includegraphics[width=0.45\textwidth]{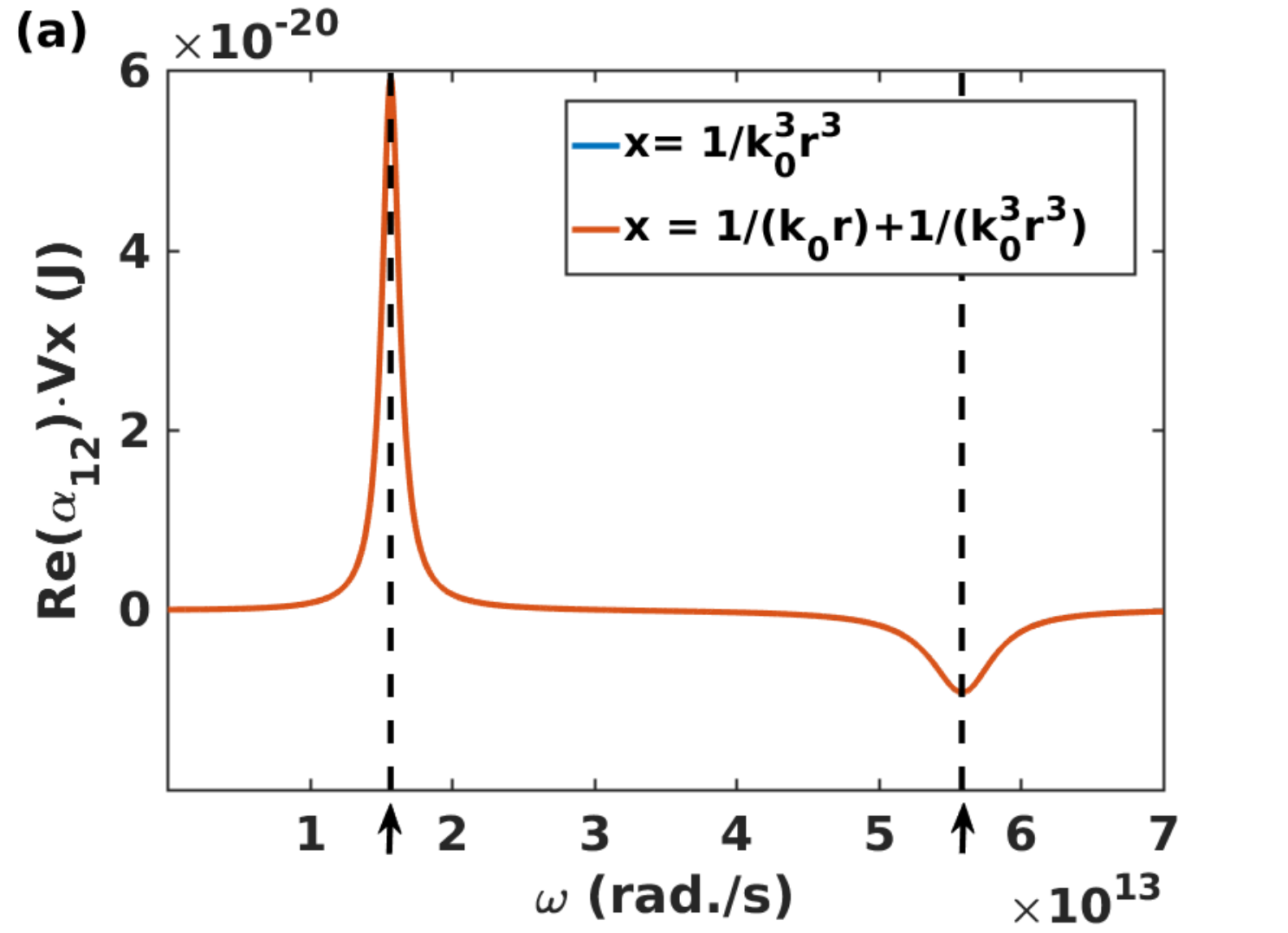}
	\includegraphics[width=0.45\textwidth]{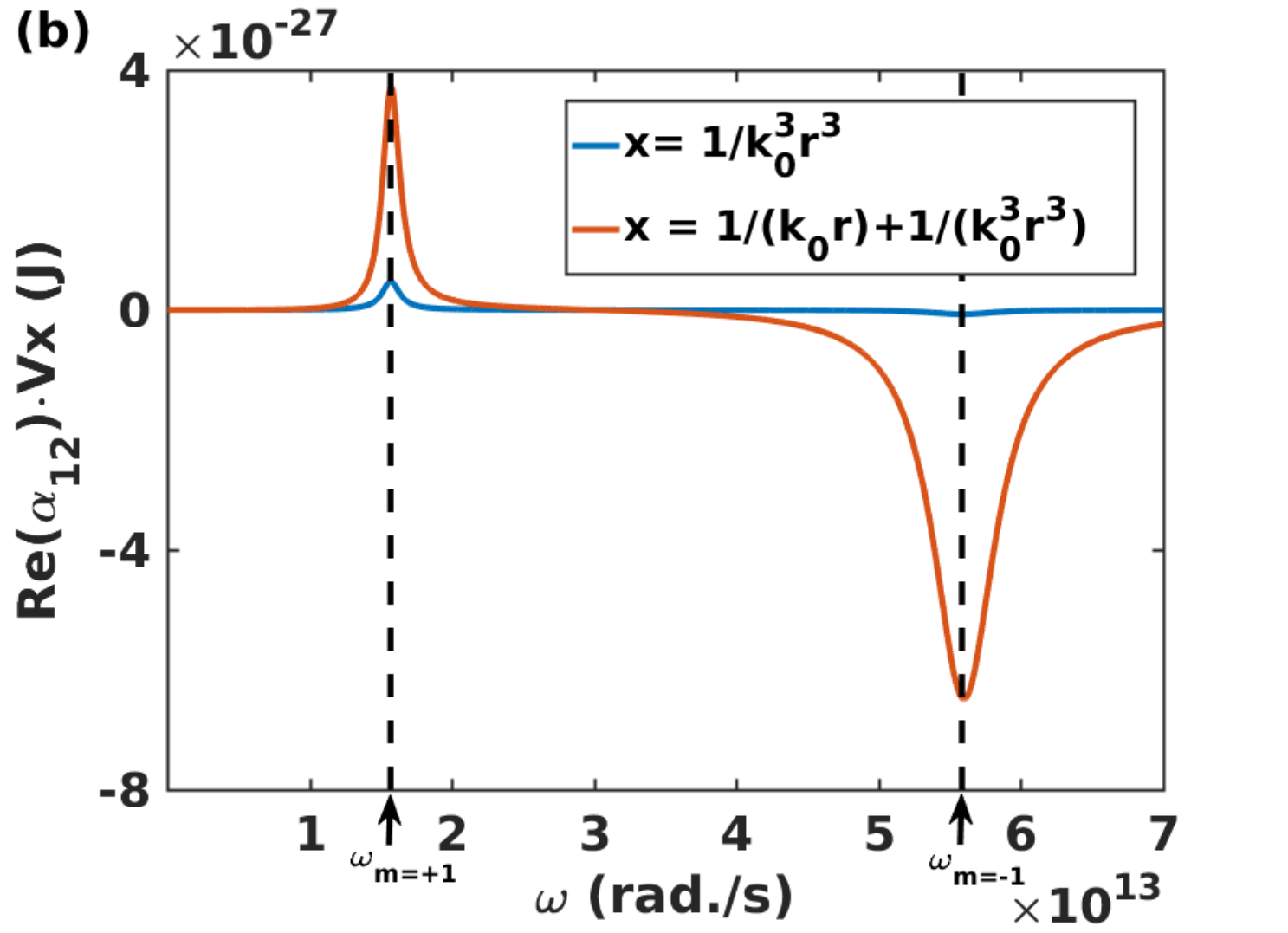}
        \includegraphics[width=0.45\textwidth]{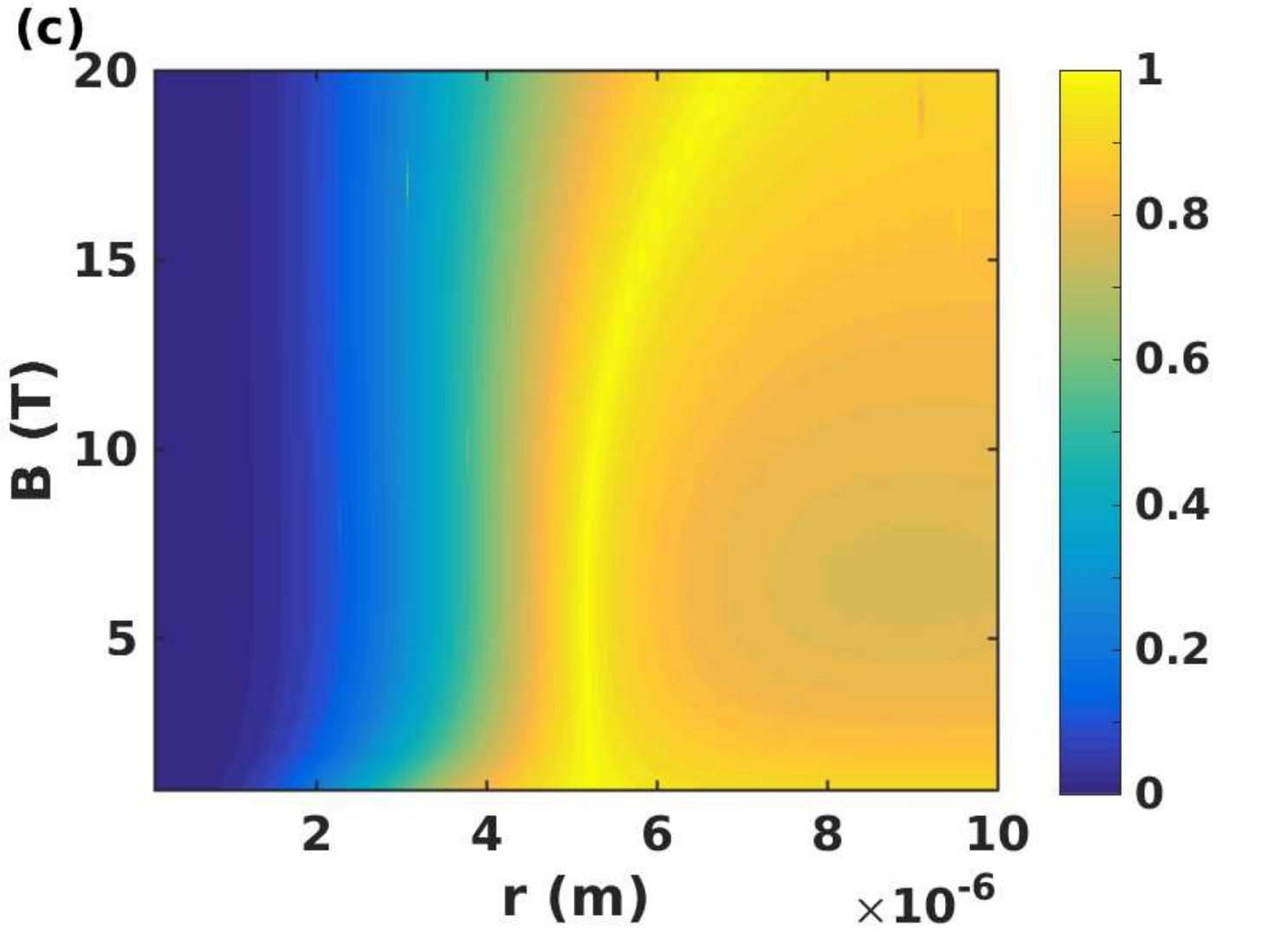}
	\caption{Plot of $\alpha_{12}' \Theta(\omega,T) k_0^3$ weighted by the factors $x = 1/(k_0 r)^3$ (near-field regime) 
                 and $x = 1 / k_0 r$ (far-field regime)  for a magnetic field of 5T at temperature $T=300 \:K$ and for $R = 100\,{\rm nm}$.
                 In (a) we have chosen a distance of $r = 100\,{\rm nm}$
                 and in (b) we have chosen a distance of $r = 50\,\mu {\rm m}$. In the near-field the low frequency mode of the $m = +1$ resonance 
                 is dominating and in the far-field regime the $m = -1$ modes dominate the spectral heat flux, so that the direction of the heat 
                 flux is changed.(c) Directionality $D$ from Eq.~(\ref{Eq:D}) as function of distance $r$ from the nanoparticle and as function of the applied magnetic field.}
	\label{alpha12}
\end{figure}

The directionality of the heat flux can now be explained in the following way:
when the field is turned off the two modes with $m = \pm1$ contribute with the same energy 
and therefore there is no net circular heat flux in this case. On the other hand, when the magnetic 
field is turned on these modes are separated in frequency. The mode with topological charge
$m = +1$ ($m = -1$) is shifted towards smaller (higher) frequencies. Correspondingly, the resonances
in $\alpha_{12}'$ are shifted. In the near-field regime we have $k_0 r \ll 1$. Then these resonances 
which appears in $S_{\omega,\varphi}$ are weighted by the factor $1/(k_0 r)^3$. Therefore in the the near-field
regime the low frequency modes with $m = -1$ are dominating. In the far-field regime $S_\varphi$ 
is weighted by the factor $1 / k_0 r$. Hence in the far-field regime the high frequency modes 
with $m = +1$ are dominating the radiative heat flux. In Fig.~\ref{alpha12}(a)-(b) we show a plot of $\alpha_{12}'$
weighted by the different factors in the near- and far-field showing how this weighting favors the contribution
of the $m = +1$ or $m = -1$ mode. This explains why we have a change of the directionality of the heat flux when 
going from the near-field to the far-field regime.

The directionality can be quantified, for example, by the quantity
\begin{equation}
  D := \frac{|S_r|}{|S_r| + |S_\varphi|}.
\label{Eq:D}
\end{equation}
If $D = 1$ we have a perfectly radial heat flux. For all other cases we have a circular component.
The smaller the value of $D$ the larger is the  $|S_\varphi|$ component. In Fig.~\ref{alpha12}(c) we show a numerical
evaluation of $D$ as function of the magnetic field and the distances for a InSb particle 100~nm radius at $T=300 K$. It can be seen that $D \approx 1$ 
for distances of about $6\,\mu{\rm m}$. This distances marks the transition between the near- and far-field
regime and therefore also the change of the direction of the circular heat flux. For larger magnetic
fields this transition is shifted towards larger distances. This is due to the fact that at $T=300 K$ the mode with $m = +1$ becomes more and more dominant when increasing the magnetic field.



\begin{figure}
\begin{center}
  \subfigure{\includegraphics[width=0.22\textwidth]{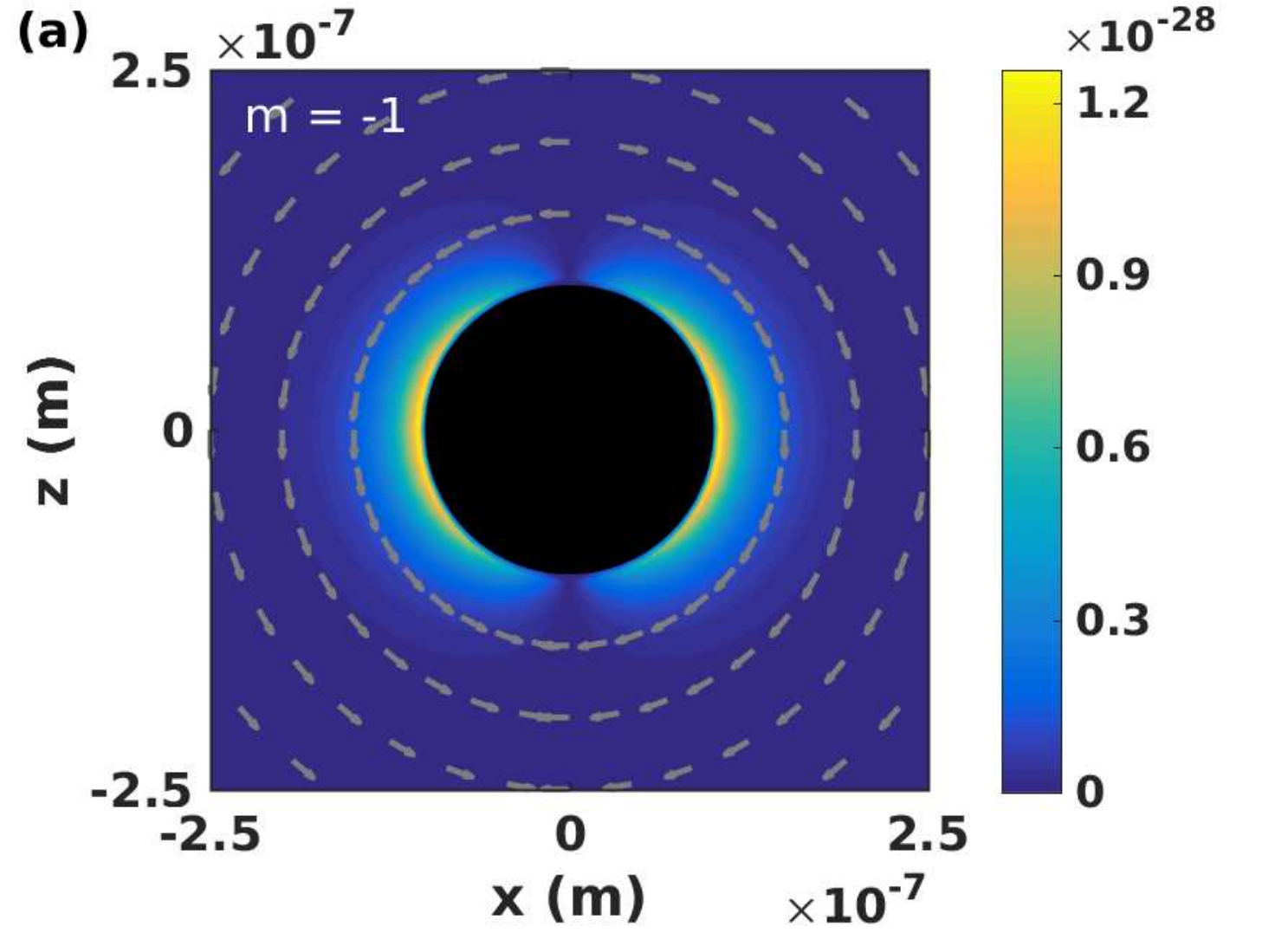}}\quad
  \subfigure{\includegraphics[width=0.22\textwidth]{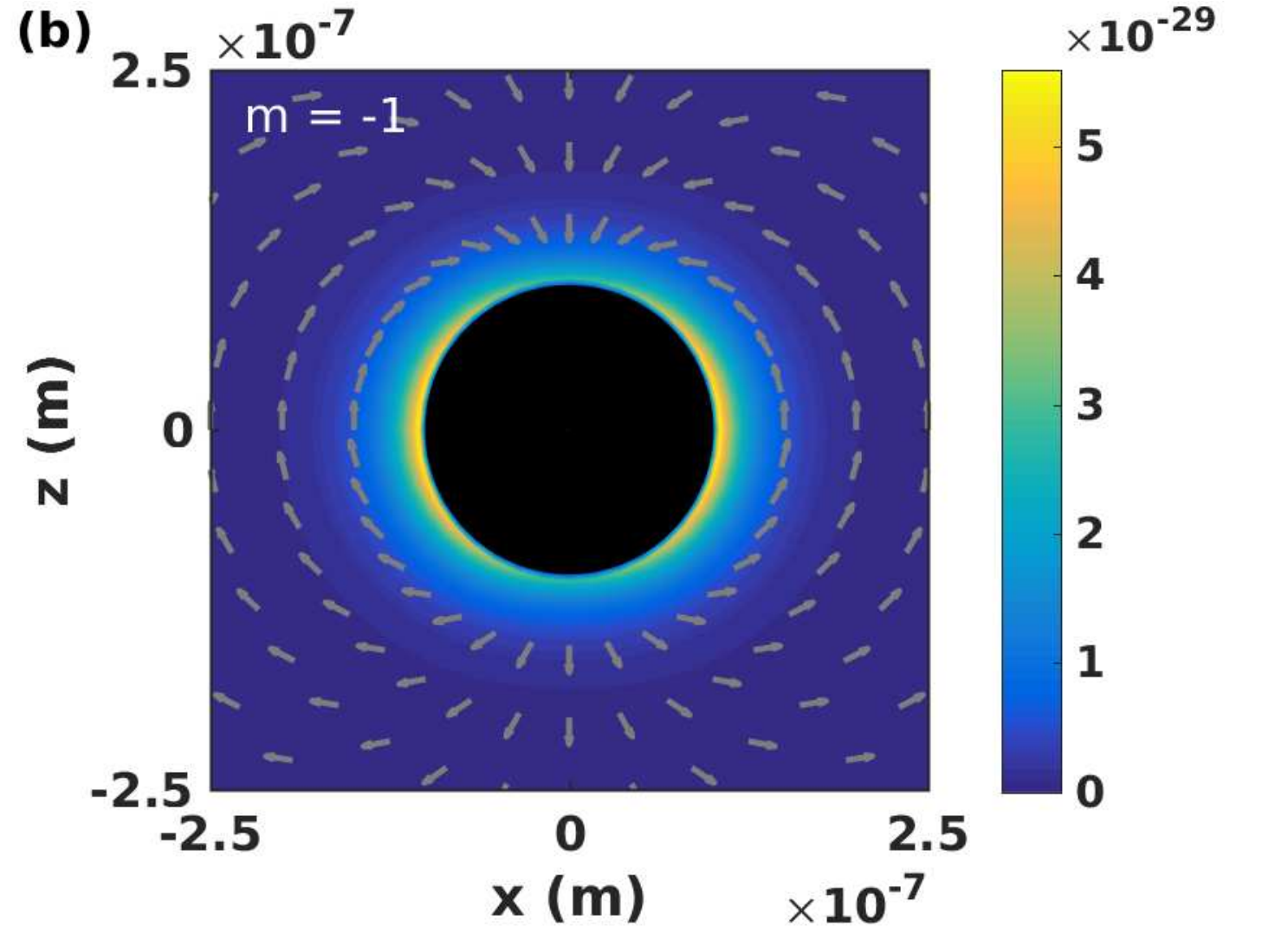}}\\
  \subfigure{\includegraphics[width=0.22\textwidth]{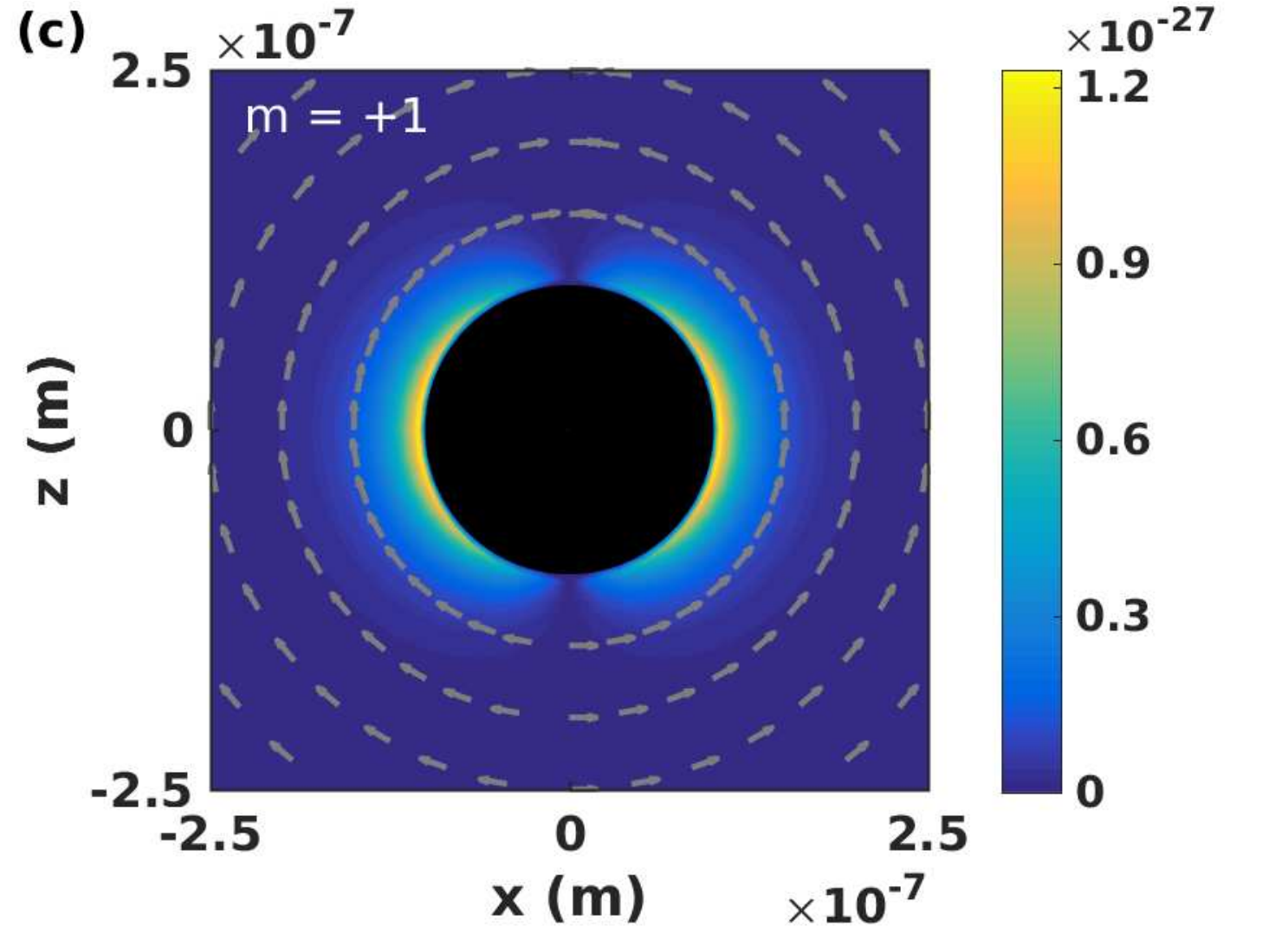}}\quad
  \subfigure{\includegraphics[width=0.22\textwidth]{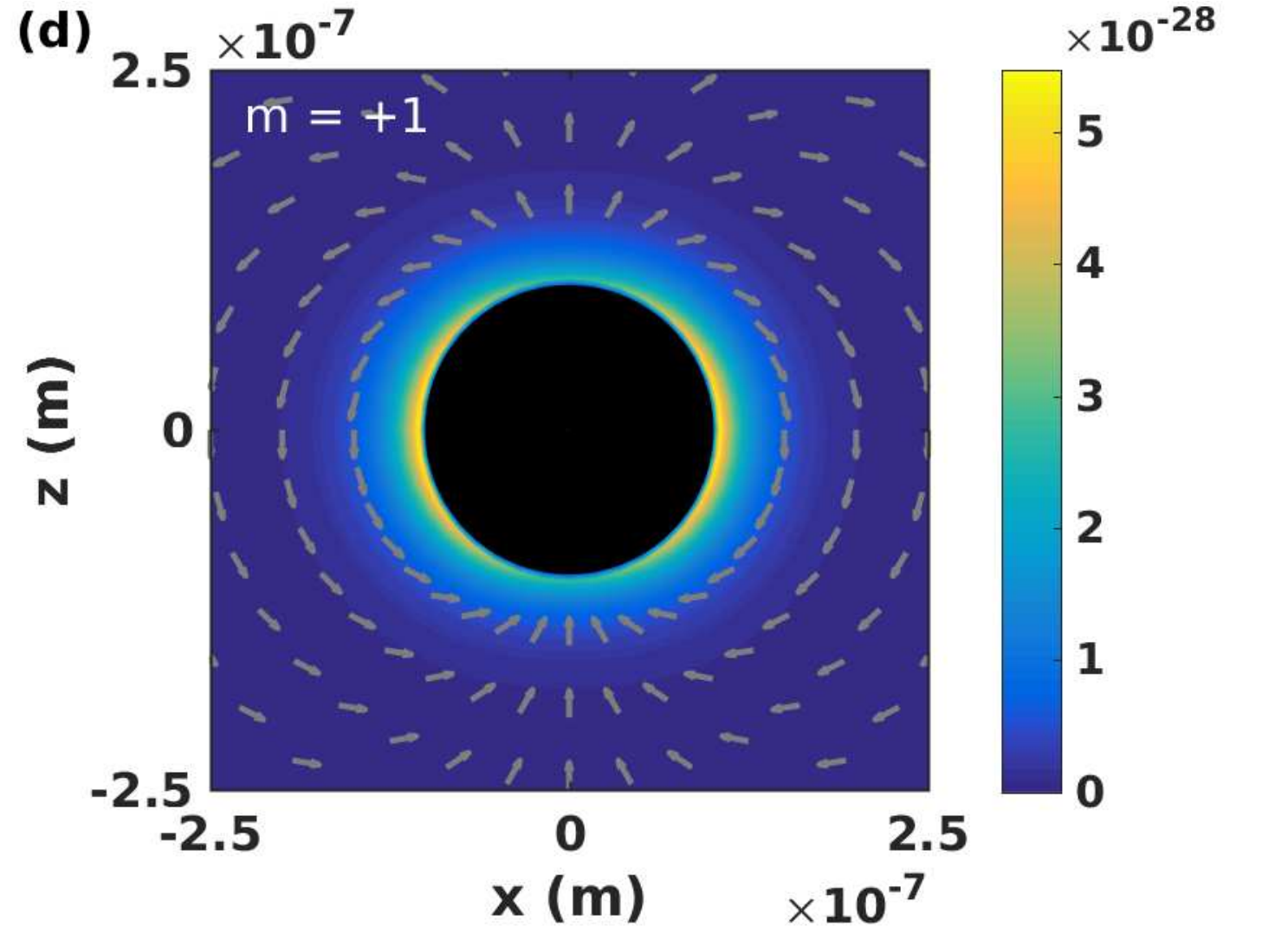}}\\
\end{center}
	\caption{Here we show the (a) spectral AM density and the (b) spectral SM density for $\omega_{m = -1}$ (top) .  The same quantities for $\omega_{m = +1}$ (bottom) are shown in (c) and (d). $B=5\:T$, $R = 100\,{\rm nm}$, and $T=300\:K$.}
	\label{resonanzen_dichten}
\end{figure}

Of course, the circular heat flux is connected to an orbital angular momentum (AM) of the electromagnetic field and also to a spin angular momentum (SM).
The total mean AM density is $\langle \mathbf{J} \rangle = \langle \mathbf{L} \rangle + \langle \mathbf{S}_{d} \rangle$,  
where the spectral SM density is defined as~\cite{Bliokh1}
\begin{equation}
  \vec{S_d}_\omega = \frac{g}{2}{\rm Im}\left(\vec{E}^*\times\vec{E}+\frac{\mu_0}{\epsilon_0}\vec{H}^*\times\vec{H}\right)
\label{spin}
\end{equation}
with $g = \epsilon_0 / \omega$ while the spectral AM density reads~\cite{Bliokh1} $ \vec{L}_\omega = \vec{r}\times\vec{P}_\omega$.
Here $\vec{P}_\omega$ denotes the canonical spectral momentum density defined  as
\begin{equation}
   \vec{P}_\omega = \frac{g}{2}{\rm Im}[\vec{E}^*(\nabla)\vec{E}+\frac{\mu_0}{\epsilon_0}\vec{H}^*(\nabla)\vec{H}], 
\label{kanimpuls}
\end{equation}
where we have adopted the same  notations as in Ref.~\cite{Bliokh1} so that $\vec{X}(\vec{Y})\vec{Z} = \sum_{i}X_i\vec{Y}Z_i$.

Similarly to the PV these quantities can straight-forwardly be evaluated.  We find that
\begin{equation}
\begin{split}
  \langle\vec{L}_\omega\rangle &= \frac{\Theta(\omega,T)k_0^4}{16\omega^2\pi^2r^2} \bigl[\frac{4}{k_0r}(\alpha_{33}''-\alpha_{11}'')\cos(\theta)\sin(\theta)\vec{e}_{\phi} \\
                               &\qquad -\frac{9+4k_0^2r^2+2k_0^4r^4}{k_0^4r^4}\alpha_{12}'\sin(\theta)\vec{e}_{\theta}\bigr].
\end{split}
\label{bahndrehimpuls} 
\end{equation}
Obviously the AM has no radial component. Furthermore, when constricting the AM onto the x-y plane by setting $\theta = \pi/2$
it becomes aligned to the z direction. Depending on the sign of $\alpha_{12}'$ it will point into the positive or negative z-direction. Since the sign of $\alpha_{12}'$ is positive for the modes with $m = +1$ and negative for the modes with $m = -1$, we obtain the an AM in the corresponding positive and negative z-direction. This is what we could already expect from the fact that both modes are connected to a clock or counterclockwise circular motion in the x-y plane.

For the spectral SM density of the thermal field emitted by the nanoparticle we find a more complicated expression given by
\begin{equation}
\begin{split}
  \langle\vec{S_d}_\omega\rangle &= \frac{\Theta(\omega,T)k_0^4}{16\omega^2\pi^2r^2}\bigl[\frac{4k_0^4r^4+2}{k_0^4r^4}\alpha_{12}'\cos(\theta)\vec{e}_r\\
                                 &\quad  +\frac{4}{k_0^4r^4}\alpha_{12}'\sin(\theta)\vec{e}_{\theta}\\
                                 &\quad +\frac{4}{k_0r}(\alpha_{11}''-\alpha_{33}'')\sin(\theta)\cos(\theta)\vec{e}_{\phi}   \bigr].
\end{split}
\label{spinerg} 
\end{equation}
It is now interesting to note that the SM density has a radial component whereas the AM has not. Therefore the SM and the AM densities
are not always showing into the same direction. Another interesting feature of the SM density is that the $\varphi$ component of the SM density is as large as 
the $\varphi$ component AM density but with opposite sign. Hence, the total AM density has no $\varphi$ component at all. 
In Fig.~\ref{resonanzen_dichten} we have plotted $\langle\vec{J}_\omega\rangle$ and $\langle\vec{L}_\omega\rangle$ for the two resonance frequencies with $m = \pm 1$. According to the relation between the AM and SM it can be seen that the AM and SM are anti-parallel for $z = 0$. For $x = 0$ the SM and AM densities are perpendicular to each other. The full integrated quantities $\langle\vec{L}\rangle$, $\langle\vec{S_d}\rangle $, and $\langle\vec{J}\rangle$ have just the same qualitative structure as the spectral quantities for $m = +1$ (not shown here). Again the $m = -1$ resonance dominates when going into the far-field regime.



To conclude, in this work we have shown that out of thermal equilibrium MO particles can be used to induce locally at nanometer scale and  in presence of an external magnetic field,  a chiral structure of heat and momentum flux. The orientation of these topological defects is driven either by the magnitude of magnetic field or by the particle temperature itself. This behavior can be traced back to the collective electronic resonances in the magnetic nanoparticle. It turns out that the degeneracy of the two dipolar resonances with topological numbers $m = \pm 1$ is lifted when a magnetic field is applied. One resonance is shifted towards smaller and the other one towards larger frequencies. As a consequence, depending on the particle temperature one of these resonances dominates the thermal radiation of the particle in its environement and this mode defines the direction of winding of flux lines. We have shown that the circular heat flux  also carries AM and SM  and demonstrated that the AM and SM point in general in different direction. They can be perpendicular or anti-parallel.
The study of these topological defects and their interaction in nanoparticle lattices remains a challenging problem. Beside its fundamental interest this effect might pave the way to promising applications in the field of thermal management at nanoscale assisted by an external magnetic field. 

The authors acknowledge financial support by the DAAD and Partenariat Hubert Curien Procope Program (project 57388963).

\end{document}